\documentclass[12pt,onecolumn]{pasj00}
\usepackage{graphicx}

\newcommand{\bm}[1]{\mbox{\boldmath $#1$}}
\SetRunningHead{Susa}{SPH with Radiation Transfer}
\title{
Smoothed Particle Hydrodynamics coupled with Radiation Transfer
}
\author{Hajime Susa}
\affil{Department of Physics, Rikkyo University, Nishi-Ikebukuro,
Toshimaku, Japan}
\email{susa@rikkyo.ac.jp}
\Received{$\langle$reception date$\rangle$}
\Accepted{$\langle$acception date$\rangle$}
\Published{$\langle$publication date$\rangle$}

\begin{document}
\maketitle
\begin{abstract}
We have been constructed a brand-new radiation hydrodynamics solver 
based upon Smoothed Particle Hydrodynamics (SPH), which works on
 parallel computer system. The code is designed to investigate the
 formation and evolution of the first generation objects at $z \gtrsim 10$,
 where the radiative feedback from various sources play important
 roles. The code can compute the fraction of chemical species e, H$^+$, H,
 H$^-$, H$_2$,  and H$_2^+$ by fully implicit time integration. It also
 can deal with multiple sources of ionizing radiation, as well as
 the radiation at Lyman-Werner band. We compare the results for a few test
 calculations with the results of one dimensional
 simulations, in which we find good agreements with each other.
 We also evaluate the speedup by parallelization, that is found to be
 almost ideal, as far as the number of sources is comparable to the
 number of processors.

\end{abstract}


\section{Introduction}
\label{intro}
One of the important objectives of cosmology today is to understand 
the way how the first generation stars or galaxies are formed and how they
affected the later structure formation, and how they reionized the
surrounding media. 
These problems have been studied intensively in the last decade, hence we
already have some knowledge on these issues. However, studies which
properly address the radiation transfer effects are still at the
beginning, in spite of their great importance.

In order to investigate the radiation transfer effects on such issues 
in realistic cosmological density field, we need the code which can deal with
hydrodynamics, gravity, chemical reactions, 
and naturally radiation transfer in
three dimension. 
So far, various kinds of numerical approaches have been taken by several authors at different levels.  
There are several codes which
focusing on the fragmentation of
primordial gas \citep{Abel00,Bromm99,Bromm02,NU99,NU01,Yoshida03}, 
without radiation transfer. 
There  also are the radiation transfer codes in which
hydrodynamics are not coupled \citep{Abel99, NUS01, Raz02, MFC03}.
Some of these codes have the merit that they can include diffuse
radiation field \citep{NUS01, Raz02, MFC03}.
Utilizing this strong point, 
these trials can give answers to the cosmic reionization problem to some
extent, however, they do not fit to investigate the problem of structure
formation because of 
the dynamical nature of the phenomena and radiative feedback effects on star/galaxy formation.
Moreover, cosmic reionization problem itself is also  coupled with
structure formation, since the formed stars, galaxies and black holes
are the sources of reionization. Therefore we need some assumptions on the
nature and number of the sources when we try to
investigate the reionization of the Universe with this type of simulations.

Another approach to these issues is suggested by
\citet{GA01}, in which paper they propose a new type of approximation,
called as Optically Thin Variable Eddington Tensor method (OTVET). They
solve the moment equations of transfer equation on the assumption 
that the closure relation is given by the Eddington tensor which is
assessed under the optically thin approximation.
The advantage of the approximation is that radiation
transfer can be coupled with hydrodynamics, since this
approximation reduces the computational cost dramatically.
Therefore they could trace the cosmic reionization history in a
self-consistent fashion.

Recently, \citet{Mellema05} and \citet{saru} have constructed fully
coupled Radiation Hydrodynamics (RHD) codes applying on the cosmic
reionization problem. 
They couple the adoptive mesh refinement hydrodynamics codes
with adoptive ray-tracing codes. They successfully trace the radiation
hydrodynamic expansion of HII region, as well as the formation of
shadows behind dense clumps.
However, their codes still do not include
H$_2$ chemistry and transfer of Lyman-Werner band radiation, although
they are
crucial for the calculation of first structure formation.

Another branch of radiation hydrodynamic approach was proposed by
\citet{KB00}, in which they show a scheme of ray tracing in Smoothed
Particle Hydrodynamics (SPH). 
SPH is often applied to the
simulations on galaxy formation and star formation, because of its
Lagrange nature and simplicity. 
They create grid points on  light rays from
sources to SPH particles. The code has an advantage that the ray tracing
is naturally incorporated into Lagrangian scheme, since they utilize the
neighbor lists of particles to find the grid points. We have parallelized
their scheme in case light rays from the source can be approximated as
parallel \citep{SU04a}. However, the parallelized code cannot be
applied to the problems with multiple sources, that are crucially
required for the first structure formation.

In this paper, we describe the newly developed RHD code by ourselves, in
which fully coupled 
hydrodynamics, gravity, chemical reactions with H$_2$ are included,
as well as 
radiation transfer of ionizing photons and the Lyman-Werner band
photons. 
One of the characteristic feature of this code
is that we employ SPH.
Radiation transfer part of the code is a natural extension
of our previous code described in \citet{SU04a}. 
Thanks to the newly developed radiation transfer and parallelization scheme, we
can include multiple sources in our new code.
Since it is a particle code, 
it also has an advantage that it has
the good affinity to GRAPE system \citep{FMK05} which can be dedicated to the
gravitational force calculations and neighbor search. 
We also show the outcomes of some calculations on the test
problems, that are compared with the results of one
dimensional simulations. 

This paper is organized as follows. We describe the detail of the code
in section \ref{code}. Then we show the results of test calculations in
section \ref{results}, and summarize in the final section \ref{summary}.

\section{Code Description}
\label{code}
In this section, we describe the basic methods of the code, dividing
into several parts. First, we describe the method to evaluate
gravitational force. Then, we explain how we treat hydrodynamics,
radiation transfer, and thermal processes in order.

\subsection{Gravity}
\label{gravity}
We use a parallel version of Barnes-Hut Tree code \citep{BH86} to assess
the acceleration owing to gravitational force. The core routine of the
tree code is originally provided by Jun Makino \footnote{http://grape.astron.s.u-tokyo.ac.jp/~makino/softwares/}, as a serial
version in FORTRAN77, which is parallelized by ourselves. 

The computational domain is decomposed by Orthogonal Recursive
Bisection (ORB) method \citep{Dubinski96}. We built up the so-called local tree and
locally essential tree, then walk along the tree structure to obtain the
interaction list. The
parallelization scheme is summarized in Figure \ref{fig:Tree}. Suppose we
have four processor units (PUs) whose computational domains
are represented by the rectangles. Now we try to assess the acceleration on
the particles in PU1. In the first step (left panel), Barns-Hut tree
structure is constructed on each PU from the particles involved in its own domain. This tree is called as a local tree, which is
not sufficient to assess the force of gravity on particles contained in
PU1. Thus we need information from other PUs. 
After making the local trees, we walk along them in other PUs
 as a ``box'' which has the same dimension as the computational domain of PU1.
 As a result, we find the tree nodes that are not opened
when they interact with the
particles in PU1 (upper middle panel). 
These tree nodes data are
sent to PU1 as the data of ``particles'', since it is not necessary to open
the nodes anymore 
when they are used to calculate the gravitational force on the
particles in PU1 (right panel).
Finally, we can reconstruct the tree structure
, i.e. locally essential tree, from
the particles in PU1 plus the ``particles''from other PUs ( lower middle
panel ). 
Utilizing this new tree structure, 
we can assess the force of gravity on the particles in PU1. We can do
the same operation on other PUs simultaneously, therefore, the acceleration
on all the particles can be evaluated. 
The tree structure is also used to obtain the neighbor SPH particle
list (see section \ref{hydro}), 
which will also be used in ray tracing routine again (see section \ref{RT}).
At present, the
code is not designed to utilize GRAPE boards, however, will be upgraded to
a parallel-GRAPE-Tree code.

\subsection{Hydrodynamics}
\label{hydro}
Hydrodynamics is calculated by SPH scheme,
based upon the method summarized in \citet{Thac00}.
The local density of the gas is evaluated by following standard
convolution method:
\begin{eqnarray}
\rho(\bm{x}_j) &=& \sum_i m_i W(|\bm{x}_i-\bm{x}_j|, h_{ij}) 
\end{eqnarray}
where
$$
h_{ij}=(h_i + h_j)/2, ~~~W(|\bm{x}|,h)=\frac{W_s(|\bm{x}|/h)}{h^3}
$$
and
\begin{eqnarray}
W_s(y) = \frac{1}{4\pi}\left\{
\begin{array}{cc}
4 - 6y^2+3y^3,~~~0\le y \le 1 &\\
(2-y)^3,~~~~~~~~~1 < y \le 2 &\\
0,~~~~~~~~~~~~~~~y >2. &
\end{array}
\right.
\end{eqnarray}
The equation of motion for each SPH particle is given by
\begin{eqnarray}
\frac{d \bm{v}_i}{dt} &=& \bm{g}_i - \sum_j m_j \left(\frac{P_i}{\rho_i^2} +
					\frac{P_j}{\rho_j^2} + \Pi_{ij}
				       \right) \times \nonumber \\
&~~&\nabla_i\left[W(|\bm{x}_i-\bm{x}_j|, h_i ) + W(|\bm{x}_i-\bm{x}_j|, h_j
	     )\right]/2 .
\end{eqnarray}
Here $\bm{g}_i$ is the gravitational acceleration and $\Pi_{ij}$ denotes
the standard 'Monaghan' artificial viscosity \citep{Monaghan92},   
\begin{eqnarray}
\Pi_{ij} &=& \frac{-\mu_{ij}{(c_i+c_j)/2}+2\mu_{ij}^2}{(\rho_i+\rho_j)/2},
\end{eqnarray}
where,
\begin{eqnarray}
\mu_{ij}=\left\{
\begin{array}{cc}
{h}_{ij}\bm{v}_{ij}\cdot\bm{x}_{ij}/(|\bm{x}_i-\bm{x}_j|^2 + 0.01{h}_{ij}^2), ~\bm{v}_{ij}\cdot\bm{x}_{ij} < 0 \\
0,~~~~~~~~~~~~~~~~~~~~~~~~~~~~~~~~~~~~~~\bm{v}_{ij}\cdot\bm{x}_{ij} \ge 0\\
\end{array}
\right.
\end{eqnarray}
The equation of motion is integrated by the standard leap-frog method.
We also employ the update scheme of SPH smoothening radius $h_i$ given in
equations (10) and (11) in \citet{Thac00}, which adjust the number of
neighbor particles to nearly constant. We set the number of neighbors
to be 100 in our simulations. We also remark that we use the tree
structure in order to obtain the neighbor list.

In order to parallelize SPH scheme,  parallelization of neighbor search
algorithm is inevitable and crucial.
The method to parallelize neighbor search algorithm is schematically
shown in Figure \ref{fig:neighbor}. Suppose we
have four PUs and try to find the neighbor list of
the particles in PU1. 
Left panel shows the first phase of the algorithm, 
in which local tree is made on each PU.
Since the neighbor particles would be included in other PUs, local tree is
not enough to find the complete list of neighbors.
In order to obtain the information from other PUs, 
the particles in the ``margin'' of PU1 are found by
 using the local trees in other PUs. The obtained data of
 particles are sent to PU1 (upper middle panel).
The ``margins''surrounding the boundary of the domain associated
 with PU1, are determined by the maximal radius among
 the particles in PU1. The radius is two times the maximal smoothening radius
 (denoted as $2h_{\rm max}({\rm PU1})$ in the upper middle panel).
Finally, the list of particles which might be in the neighbor of
 particles in PU1 is completed (right panel). Using this list, locally
 essential tree is reconstructed, which is used to find neighbor
 particles in PU1 (lower left panel).
We can perform same operations on other PUs simultaneously, hence we
obtain the complete list of neighbor particles for all particles.

\subsection{Ray Tracing}
\label{RT}
It would be possible to solve full three dimensional radiation hydrodynamic
problem in the next decade, but it is not possible at present. Actually,
\citet{NUS01} have performed fully three dimensional simulation of  
radiation transfer for ionizing photons including the effects of diffuse
radiation. However, the computational cost was so expensive that it
cannot be used directly for the radiation hydrodynamic problems.

In our code, so called on-the-spot approximation \citep{Spitzer78} is employed. 
We solve the transfer of ionizing photons directly from the
source, whereas we do not solve the transfer of diffuse photons. 
Instead, it is
assumed that the recombination photons are absorbed in the very
neighbour of the spatial position where they are emitted.
Thus, the radiation transfer equation for intensity $I_{\nu}$ at
frequency $\nu$ reduces to
a very simple equation:
\begin{eqnarray}
\frac{d I_\nu}{dr}&=& -n_{\rm HI}\sigma_\nu I_\nu,
\label{eq:radtr}
\end{eqnarray}
where the re-emission term is not included, since it is taken into
account in the reduced recombination rate. This transfer equation (\ref{eq:radtr}) itself is easily
integrated to find $I_\nu \propto \exp{\left(-\tau_\nu\right)}$, where
the optical depth $\tau_\nu$ is defined as:
$$
\tau_\nu \equiv \int_0^r n_{\rm HI}\sigma_\nu dr,
$$
where $r$ is the distance from the source, $n_{\rm HI}$ is the HI
density, and $\sigma_\nu$ denotes the photoionization cross section. 
The on-the-spot approximation reduces the computational
cost drastically, if the number of the sources is much smaller than the
number of SPH particles.
We also note that the frequency dependence of optical depth is originated
from $\sigma_\nu$, which dependence is already known.
Therefore, all we need is the optical depth at a certain frequency 
from in order to assess the optical depth at arbitrary frequency. 

It is not trivial to assess the optical depth in SPH
scheme. \citet{KB00} have proposed a scheme for SPH, which utilize the
neighbour lists of SPH particles to create grid points on the rays from
the source to SPH particles. In fact, we have parallelized the scheme to
use in our
previous numerical simulations on the photoevaporation of proto-dwarf galaxies\citep{SU04a,SU04b}.
However, in case the number of SPH particles are very large, computational cost of grid points
creation on the light ray is not negligible. Moreover, the parallelized
version of the code cannot be applied to the problems including multiple sources.
Thus, we use the similar
method which is slightly different from the original Kessel-Deynet's scheme.

In our new scheme, we do not create so many grid points on the light
ray. 
Instead, we just create one grid point per one SPH particle in its neighbor.
The method is described schematically in Figure \ref{fig1}. 
Suppose that we are trying
to assess the optical depth of the particle ``0'' from two sources ``S1''
($\tau_{0\leftarrow S1}$) and ``S2''($\tau_{0 \leftarrow S2}$). As the first step, we make a
thorough search on the neighbor list of the particle
``0''to find the ``upstream'' particles 
which are closest to the light rays.  
The ``upstream'' particle are found by a simple criterion, that the angles
$\theta_1$ and $\theta_2$ are the smallest.
In the case of Figure \ref{fig1}, particles ``4'' and ``2'' are
chosen as the ``upstream''particles.
 
Next, we create the grids on the rays. 
The grid points on the rays are
found as the intersections (marked as wedge-shapes in the figure) of
the light rays and the sphere with the radius which equals to the
distance between the sources and the ``upstream'' particles.
The physical variables on the particles are mapped onto the
intersection points from the ``upstream'' particles. 
Using the mapped quantities, 
the differential optical depths between the intersections
 and particle ``0'' 
($\Delta\tau_{0\leftarrow S1}$ and $\Delta\tau_{0\leftarrow S2}$) 
are evaluated by simple trapezium formula. 
Finally, we obtain the optical
depths as
\begin{eqnarray}
\tau_{0 \leftarrow S1} &=& \tau_{4 \leftarrow S1} + \Delta\tau_{0\leftarrow S1}\\
\tau_{0 \leftarrow S2} &=& \tau_{2 \leftarrow S2}+ \Delta\tau_{0\leftarrow S2}
\end{eqnarray}
where $\tau_{4 \leftarrow S1}$ and $\tau_{2 \leftarrow S2}$ denote the optical depth form S1 to
particle 4 and the one from S2 to particle 2. Clearly, we need the optical
depth of particles 4 and 2 before we assess $\tau_{0 \leftarrow S1}$ and
$\tau_{0 \leftarrow S2}$. Therefore, we have to evaluate the optical depth of all
particles in reverse order of the distance from the source.

If we try to parallelize the code, we need communications of optical depth
between the processors, since we have to assess the optical depth of
each particles in order in the computational domain decomposed by ORB(
see section \ref{gravity}). Consequently, we have to wait to assess the
optical depth until the information from ``upstream'' processors arrive. 
Because of this waiting time, the parallelization efficiency becomes
very poor.

In order to overcome this problem, we incorporate
two tricks.
First, we use Multiple Wave Front (MWF) technique which was originally
developed by \citet{NUS01}. Following their method, we can overlap the
calculations of optical depth for multiple sources. 
Since the number of ``waiting''
processors is reduced for multiple sources, the parallelization
efficiency becomes the better the number of sources gets larger.
Second, we use the scheme proposed by \citet{Hienemann05}, which
accelerates MWF for small number of sources. In their method, processors
have to wait only while the ``upstream'' processors are calculating the
optical depth of particles on the boundary of computational
domain( See Figure \ref{fig2}). 
Therefore, the ``waiting'' time is greatly reduced, which also
helps to improve the parallelization efficiency.

Figure \ref{fig2} shows an example of propagation of
transactions in computational domains. Suppose we have two sources
(denoted by two stars) and four PUs. 
Each panel shows a step of transaction,
which is composed of calculations and communications. 
In the shaded area, the optical depths of SPH particles are calculated, and the
arrows between the rectangles show the required communications. We have
5 steps in this example, which guarantees that the information from
upstream PUs have arrived before the calculations of optical depths
in its own processor. The PUs marked by numbers \fbox{1} and/or \fbox{2} denote
the one in which the calculation for  sources \fbox{1} and/or \fbox{2}
have finished. 
In the first panel, all four PUs are calculating
the optical depth for both of the sources simultaneously, 
assuming the optical depths at the boundary particles are zero. 
The parallelization efficiency is very close to unity in this first step. 
In the second panel, the calculated optical depths are sent to the
neighbor downstream PUs. In the third panel, the optical depths at the
opposite side of the boundary are assessed by just summing up the received
optical depths at the boundary and the optical depths calculated at the
first step in the attached PUs. The calculated boundary data are sent to the downstream again
(fourth panel). After receiving all the required data at the boundary,
the optical depths in the inner part are calculated by the sum of the 
received boundary value and the optical depth evaluated at the first step.

Here we have to remark that this parallelization scheme could introduce
very slight difference among the numerical results of various number of
processors. In the above scheme, we have to give orders among the
processors, otherwise we cannot define the ``upstream'' or
``downstream'' processors. This means when we search the ``upstream
particle'' for an SPH particle, one cannot choose the particles located in
``downstream'' processors.
Consequently, especially in case both of the source and the SPH particle
under consideration are located near
the boundary of the processors, such ordering constraint could give
different choice of the ``upstream'' particle from the one obtained in serial calculation,
since there are no such additional constraints for the serial case.
Thus the numerical results obtained from 
different number of processors are not exactly the same.

\subsection{Chemical Reactions and Energy Equations}
The non-equilibrium chemistry and radiative cooling 
for primordial gas are calculated by the code
developed by \citet{SuKi00}, in which H$_2$ cooling and
reaction rates are taken from \citet{GP98}. Atomic cooling by hydrogen
is also taken into account, utilizing the fitting formula given in
\citet{FK94}.
Helium is not included in our code at present, but it will be in the
near future {(see Appendix \ref{app})}.
Since we are primarily interested in the formation of first objects and
their feedback processes, we neglect the cooling by metals and dusts 
in the present code. We take into consideration the chemistry for H, H$^+$, H$^-$, H$_2$, H$_2^+$ and e$^-$.

The photoionization rate of HI and the photoheating rate for each SPH
particle labelled as $i$ are 
respectively given by 
\begin{eqnarray}
k_{\rm ion}^{(1)}(i)&=& n_{\rm HI}(i)\int_{\nu_L}^\infty\!\!\!\!\int
 \frac{I_\nu\left( i \right)}{h\nu} \sigma_{\nu} d\Omega d\nu, \label{eq:rateion} \\
\Gamma_{\rm ion}^{(1)}(i)&=& n_{\rm HI}(i)\int_{\nu_L}^\infty
 \!\!\!\!\int \frac{I_\nu\left( i \right)}{h\nu} \sigma_{\nu} (h\nu-h\nu_L)d\Omega d\nu. \label{eq:rateheat}
\end{eqnarray}
Here $n_{\rm HI}(i)$ represents the number density 
of neutral hydrogen at $i$-th particle, $\sigma_{\nu}$ is
the photoionization cross section, which is taken from the table in \citet{SK87}.
The frequency at the Lyman limit is denoted by $\nu_{\rm L}$, and $\Omega$ is the solid angle. $I_{\nu}(i)$ is
the intensity of the ultraviolet radiation that irradiate $i$-th
particle, which is obtained by the scheme described in section \ref{RT}.
Remark that the equations (\ref{eq:rateion}) and (\ref{eq:rateheat})
denote the general formulae of those rates at arbitrary spatial
position, omitting the suffix $i$.

In case the optical depth for ionizing photon for a single SPH particle is moderate, the equations
(\ref{eq:rateion}) and (\ref{eq:rateheat}) are valid. If the
optical depth becomes much larger than unity, however, those expressions
could lead to essentially zero ionization and heating rates because the
equations do not conserve the number of photons numerically. 
Thus, we need something like photon conserving method \citep{KB00,Abel99}
in order to avoid this difficulty. 

For the purpose to find better formulae for optically thick regime, we
combine equations (\ref{eq:rateion}),(\ref{eq:rateheat}) without the
suffix $i$ and (\ref{eq:radtr}) to rewrite the ionization rate and the
photoheating rate as follows:
\begin{eqnarray}
k_{\rm ion}&=& -\frac{1}{4\pi r^2}\frac{d}{dr}\int_{\nu_L}^\infty \frac{L_\nu^*\exp\left(-\tau_\nu\right)}{h\nu} d\nu, \\
\label{eq:rateion2}
\Gamma_{\rm ion}&=&-\frac{1}{4\pi r^2}\frac{d}{dr}\int_{\nu_L}^\infty \frac{L_\nu^*\exp\left(-\tau_\nu\right)}{h\nu}(h\nu-h\nu_L)d\nu,
\label{eq:rateheat2}
\end{eqnarray}
where $L_\nu^*$ denotes the intrinsic luminosity of the source and 
$r$ is the distance from the source.
Here we also use the formal solution of equation (\ref{eq:radtr}) and
assume that size of the source is much smaller than $r$.

Now we are ready to derive the better formula, that are the ``volume
averaged'' rates:
\begin{eqnarray}
k_{\rm ion}^{(2)}(i)&\equiv& \overline{k_{\rm ion}} = \frac{\displaystyle\int_{r_i-\Delta r_i/2}^{r_i+\Delta
 r_i/2}k_{\rm ion}r^2dr}{\displaystyle\int_{r_i-\Delta r_i/2}^{r_i+\Delta
 r_i/2}r^2 dr},\nonumber \\
&=& \frac{3}{\Delta r_i} \frac{\Phi_1(r_i-\Delta
 r_i/2)-\Phi_1(r_i + \Delta r_i/2)}{3r_i^2 + \Delta r_i^2/4},\label{eq:rateion3}\\
\Gamma_{\rm ion}^{(2)}(i)&\equiv& \overline{\Gamma_{\rm ion}} = \frac{\displaystyle\int_{r_i-\Delta r_i/2}^{r_i+\Delta
 r_i/2}\Gamma_{\rm ion}r^2dr}{\displaystyle\int_{r_i-\Delta r_i/2}^{r_i+\Delta
 r_i/2}r^2 dr},\nonumber \\
&=&\frac{3}{\Delta r_i} \frac{\Phi_2(r_i-\Delta
 r_i/2)-\Phi_2(r_i + \Delta r_i/2)}{3r_i^2 + \Delta r_i^2/4},\label{eq:rateheat3}
\end{eqnarray}

where
\begin{eqnarray}
\Phi_1(r)&=&\int_{\nu_L}^\infty \frac{L_\nu^*}{4\pi}\frac{\exp\left(-\tau_\nu\right)}{h\nu} d\nu,\label{eq:numion}\\
\Phi_2(r)&=&\int_{\nu_L}^\infty
 \frac{L_\nu^*}{4\pi}\frac{\exp\left(-\tau_\nu\right)}{h\nu}(h\nu-h\nu_L) d\nu \label{eq:numheat}.
\end{eqnarray}

Here $r_i$ is the distance between the source and $i$-th particle, $\Delta r_i$ denotes the spatial step of ray tracing integration,
which satisfy 
\begin{eqnarray}
r_j = r_i - \Delta r_i /2, 
\end{eqnarray} 
where $j$ denotes the label of the particle located at ``upstream'' of
$i$-th particle, which is found by the algorithm described in section \ref{RT}.

The above formula has an important advantage 
that the propagation of ionization front is
properly traced even for a large particle separation with 
optical depth greater than unity. However, in the neighbor
of the source, equations (\ref{eq:rateion}) and (\ref{eq:rateheat}) might give
better value than equations (\ref{eq:rateion3}) and
(\ref{eq:rateheat3}) do because of the poor spatial resolution and high
ionization degree. 
Thus, we employ following formula for ionization and
photoheating rates, in which the rates are switched to one another
depending on the local optical depth:
\begin{eqnarray}
k_{\rm ion}(i)&=&
\frac{k_{\rm ion}^{(1)}(i)}{1+\Delta\tau_i/\Delta\tau_{\rm cr}} + 
\frac{\Delta\tau_i/\Delta\tau_{\rm cr} k_{\rm ion}^{(2)}(i)}{1+\Delta\tau_i/\Delta\tau_{\rm cr}},\\ \label{eq:rateion4}
\Gamma_{\rm ion}(i)&=& 
\frac{\Gamma_{\rm ion}^{(1)}(i)}{1+\Delta\tau_i/\Delta\tau_{\rm cr}} + 
\frac{\Delta\tau_i/\Delta\tau_{\rm cr}\Gamma_{\rm ion}^{(2)}(i)}{1+\Delta\tau_i/\Delta\tau_{\rm cr}}.\label{eq:rateheat4} 
\end{eqnarray}
Here $\Delta\tau_{\rm cr}$ is a constant at which
the switching occurs, that is $10^{-3}$ in our simulations. $\Delta\tau_i$ is the local optical depth defined as:
$$
\Delta\tau_i \equiv \frac{1}{2} \left(n_{\rm HI}(j) + n_{\rm HI}(i)\right) \sigma_{\nu_L}\Delta r_i.
$$
We also remark that the integral in frequency space in
equations (\ref{eq:numion}) and (\ref{eq:numheat}) can be performed in
advance of the simulations.
As a result, $\Phi_1$ and $\Phi_2$ are found to be the functions of the optical depth only
at the Lyman limit, since we already know the frequency dependence of
the cross section $\tau_\nu$, which is proportional to $\sigma_\nu$. 
Therefore we can create tables of
$\Phi_1$ and $\Phi_2$ as functions of $\tau_{\nu_L}$,  or we have
analytic formula for some particular cases \citep{SU00,SuKi00,NUS01}. 
Consequently, we need that optical depth
just at the Lyman-limit by ray tracing.
{We also note that this formulation for pure hydrogen gas can be
extended to the hydrogen + helium gas ( Appendix \ref{app} ), although
it has not been implemented yet.}

The H$_2$ photodissociation rate is also evaluated by similar method. 
In order to assess the photodissociation rate, the transfer equation of
Lyman-Werner(LW) radiation is necessary. However, the frequency dependent
line transfer of LW band takes too much computational cost. Thus, we assume
the ``self-shielding function'' derived by \citet{DB96}, which gives the
photodissociation rate of H$_2$ as
\begin{equation}
k_{\rm dis}^{(1)} = k_{\rm dis}^{\rm (0)}
 \left(\frac{r_0}{r}\right)^2f_{\rm sh}\left(\frac{N_{\rm H_2}}{10^{14}{\rm cm^{-2}}}\right)
\label{eq:dissociation1}
\end{equation}
where
$$
f_{\rm sh}(x) = \left\{
\begin{array}{cc}
1,~~~~~~~~~~~~~~x \le 1 &\\
x^{-3/4},~~~~~~~~~x > 1 &
\end{array}
\right.
$$
Here $N_{\rm H_2}$ is the column density of hydrogen molecules measured
from the source star, $k_{\rm dis}^{(0)}$ is the unshielded
photodissociation rate in the neighbor of the source at $r = r_0$. 
We also multiply the geometrical dimming factor, $(r_0/r)^2$. 
The simplest method is to use equation
(\ref{eq:dissociation1}) directly, which works well in the case $\Delta
N_{\rm H_2} \lesssim 10^{14}{\rm cm^{-2}}$ is satisfied. Here  $\Delta
N_{\rm H_2}$ denotes the H$_2$ column density of a single spatial grid generated along the light ray, i.e. $\Delta N_{\rm H_2}= \frac{1}{2} \left(n_{\rm H_2}(j) + n_{\rm H_2}(i)\right) \Delta r_i$.

However, for $\Delta N_{\rm H_2} \gg 10^{14}{\rm cm^{-2}}$, the
photodissociation rate is significantly reduced by the absorption in LW
band while LW photons are passing through only a single grid. This could delay the propagation of
H$_2$ dissociation front because the significant amount of dissociation
photons are absorbed in a single grid without being used to dissociate
H$_2$. In order to overcome this
problem, we take similar approach to that in photoionization problem. In
the case of photoionization, we use a sort of 'photon conserving' method
by integrating the photoionization rate over a single grid. Similarly, we
integrate the photodissociation rate over a single grid, and assess the
volume averaged photodissociation rate as follows:

\begin{eqnarray}
\hspace{-5mm}&&k_{\rm dis}^{(2)}(i)\equiv \overline{k_{\rm dis}} = \frac{\displaystyle\int_{r_i-\Delta r_i/2}^{r_i+\Delta
 r_i/2}k_{\rm dis}r^2dr}{\displaystyle\int_{r_i-\Delta r_i/2}^{r_i+\Delta
 r_i/2}r^2 dr}\nonumber \\
 \hspace{-5mm}=&& \frac{r_0^2}{r_i^2 + \frac{\Delta
  r_i^2}{12}}\frac{k_{\rm dis}^{(0)}}{\Delta r_i n_{\rm H_2}(i)} \int_{N_{\rm
 H_2}^-}^{N_{\rm H_2}^+} f_{\rm sh}\left(\frac{N_{\rm H_2}}{10^{14}{\rm cm^{-2}}}\right)dN_{\rm H_2}
\label{eq:dissociation2}
\end{eqnarray}
where $N_{\rm H_2}^\pm$ denote the H$_2$ column density at $r_i\pm\Delta
 r_i/2$, respectively. Note that the factor ${k_{\rm
 dis}^{(0)}}/{n_{\rm H_2}}$ can be factorized outside the integral since
 $k_{\rm dis}^{(0)}$ is proportional to $n_{\rm H_2}$. 
The last integral in the above equation is
 easily performed analytically. Thus, we obtain the expression of volume
 averaged photodissociation rate in our scheme. Following the method
 in the case of photoionization, we also employ the switching scheme from/to
 the optically thin to/from optically thick regime described in equation
 (\ref{eq:rateion4})~(switches from/to expression in equation
 (\ref{eq:dissociation1}) to/from equation
 (\ref{eq:dissociation2})).

Now we are ready to integrate the
rate equations and energy equations implicitly:

\begin{eqnarray}
y_i^{(n+1)}&=& y_i^{(n)} + k_i^{(n+1)}\Delta t, \\
\epsilon^{(n+1)}&=& \epsilon^{(n)} + \left(\Gamma^{(n+1)} -
				      \Lambda^{(n+1)}\right)\Delta t +
 \Delta t\left[\frac{d\epsilon}{dt}\right]_{\rm ad}^{(n)}.
\end{eqnarray}
where
\begin{eqnarray}
\left[\frac{d\epsilon}{dt}\right]_{\rm ad}^{(n)} = \sum_j
 m_j \left(\frac{P_i}{\rho_i^2} +\frac{1}{2}\Pi_{ij}\right)\left(\bm{v}_i-\bm{v}_j\right)
\nabla_i W(|\bm{x}_i-\bm{x}_j|, h_{ij} )
\end{eqnarray}

Here the suffix on the right shoulders of the characters denote the labels
of the grid in time coordinate. $\Delta t$ is the time step, which is
given by the algorithm discussed in section \ref{timestep}.
$y_i$ and $\epsilon$ represent the
$i$-th chemical species and internal energy of the gas. $k_i$, $\Gamma$
and $\Lambda$ denote the reaction rate of $i$-th species, photoheating
rate, and cooling rate, all of which are assessed at ($n+1$)-th time step,
while the last term in the second equation denotes the adiabatic term, which
is evaluated at $n$-th time step. In other words, only the adiabatic
term is taken into account in an explicit form. Using these equations, 
we find the
solutions $\{y_i^{(n+1)}\}$ and $\epsilon^{(n+1)}$ for all particles
by iterative operations.

\subsection{Time Step Control}
\label{timestep}
It is possible to
use individual time step control depending on the local density in
hydrodynamic simulations. However, in radiation hydrodynamic
simulations, the justification of the scheme is unknown and uncertain
since the effects of radiation field is quite global.  Therefore, we
use synchronous time steps for all particles.

As a first step, $\Delta t$ is set as the minimal value among the Courant
conditions and the time scales of accelerations for all SPH particles:
\begin{eqnarray}
\Delta t_{\rm hydro} = {\rm min}\left[\frac{h_i}{{\rm max}(c_{{\rm s}i},
			v_i)}, \sqrt{\frac{h_i}{\dot{v_{i}}}} \right]~~~~~{\rm for
			~all~ {\it i}}.
\end{eqnarray}

Afterwards, we try to find solutions of the rate equations and energy
equations for all particles 
by the implicit time integration solver. 
If we fail to find the
solutions, i.e. the convergence is very slow, then we divide the time
step by two, and try to find the solutions again. We continue this operation
until we find the solutions within reasonable iteration steps (nominally
20 ). 
After finding the solution for energy equations and chemical reaction equations, we update the particle positions and
velocity by explicit leap-flog time integration.

\section{Results of Test Calculations}
\label{results}
We perform various tests with the code, and will show the results 
comparing with the numerical solutions from 
one dimensional radiation hydrodynamic simulations. 
We use 8 nodes ( 16 Xeon processors with gigabit ethernet) and 
1048576 SPH particles for the tests unless stated. 
\subsection{Expanding HII Region in Uniform Media}
In this test, we put a single source at the center of the uniform gas cloud
and trace the propagation of
ionization front.
The initial number density of the gas is 
$n_{\rm H} = 0.01 {\rm cm^{-3}}$,
and the temperature is $T=3\times 10^2{\rm K}$.
{The ionizing photon luminosity of the source is $S=1.33\times 10^{50} {\rm
s^{-1}}$ and the spectrum is black
body with $T_* = 9.92\times 10^4$K, that are the parameters of POPIII
stars with 120$M_\odot$ \citep{Baraffe}.}

Figure \ref{fig:w0if} shows the propagation of ionization front 
($r_{\rm I}$ ).
Numerical results from the code are shown by red dots, whereas the dotted line 
denotes the results from the one dimensional Piecewise Parabolic Method (PPM) code \citep{CW84}. 
The results agree well with each other. Note that 1D-PPM code is also
tested by the comparison with 
analytic solutions and results from the similar code
developed by Tetsu Kitayama \citep{Kitayama04}, and the two results agree
very well with each other.

Figure \ref{fig:w0} shows the spatial distributions of physical quantities (HI
fraction , temperature, density, H$_2$ fraction ) at different snapshots
( $t = 10^7{\rm yr},10^8{\rm yr},10^9{\rm yr}$). 
The results are also compared with the plots of one dimensional simulation.
Slight scatter and deviation
is found in the spatial distribution, especially at later epoch. 
However, the agreement between the two results are acceptable.

\subsection{Expanding HII Region in $\rho\propto r^{-2}$ Density Profile}
\label{coreenv}
In this test, density distribution is different from the previous
test. We assume ``core-envelope'' structure, where the uniform central
core density 
$n_{\rm H} = 100 {\rm cm^{-3}}$ whose radius is $10$pc.
We also assume the density of the outer envelope that decline 
as $\propto r^{-2}$, which is typical of the slope in collapsing
prestellar gas cloud. 
Since the Sr\"omgren radius in the core is slightly smaller than 
the core radius, initial R-type ionization front is stopped in the core.
Afterwards, the ionization front changes to D-type.
While the D-type front propagating in the envelope, the front type
becomes R-type again, because the the slope is steeper than the
critical case where $\rho \propto r^{-1.5}$\citep{Franco90}.

The results are compared with the 1-D results again in Figures.\ref{fig:w2if} and \ref{fig:w2}. 
Three different lines shown in Figure \ref{fig:w2} correspond to
$t = 4\times 10^5{\rm yr},10^6{\rm yr},4\times 10^6{\rm yr}$,
respectively. 
At the final epoch, the type of the ionization front is
being changed to R-type from D-type.
{Again, we find good agreement between the two results, although we have
slight difference at the final output. 
At the final epoch, the propagation of the shock front is
slightly delayed, which might be cased by the error in evaluating the optical
depth in front of the density peak. In fact, the width of the
density peak at the D-front is broader than that of the 1D
result. Consequently, the ionized region could be slightly
smaller in SPH simulation, which can introduce the delay of propagation time of the shock front. }

\subsection{Expanding HII Region with Multiple Sources}
\label{multi_source}
We also show the HI and density structures that are
created by multiple sources. We perform RHD simulations with randomly
distributed ten sources in the uniform spherical cloud~($n_{\rm
H} = 1 {\rm cm^{-3}}$, R=$1$ kpc). 
All of the sources are assumed to have the same
effective temperature and the luminosity as in the previous test.
The number of particles used in this simulation is half of the previous
two tests, i.e. $N_{\rm SPH} = 524288$.
 
The panels in Figure \ref{fig:multi_test} show the time evolution of the
density (left column) and HI fraction (right column) distributions on a
certain spatial slice {of the cubic region that are hollowed out
from the spherical simulated cloud}.

In the first phase, all HII regions
expand as R-type and and they reach Str\"omgren radius (from top panels
to middle panels). Afterwards,
the photoheated gas starts to expand dynamically, and dense shells are
formed surrounding the sources. The shells collide with each other, and
web of dense shells is formed (bottom panels). 
We use this calculation as the benchmark of our code in the next
section.

\subsection{Speedup by parallelization}
We assess the speedup of parallelization by performing simulations
with multiple sources in uniform media described in section \ref{multi_source}.
The initial setup of the
simulations are the same as in section \ref{multi_source}. 
We use 2, 4, 8, 16 and 32
processors (1, 2, 4, 8 and 16 nodes) for the same setup, 
and compare the clock time per single time step with each other. 

We use the first model of FIRST cluster (consists of 16 nodes with 32 processors) in University of Tsukuba for
these test runs.
The newly being developed PC cluster
``FIRST'' is designed to elucidate the origin of first generation objects
in the Universe through large-scale simulations.
The cluster will consist of 256 nodes 
( each node has dual Xeon processors ) 
connected by two dimensional
hyper-crossbar quad-gigabit network utilizing trunk technology.  
The most
striking feature of the system is that each node has small GRAPE6
PCI-X board called Brade-GRAPE. The Brade-GRAPE has four GRAPE6 chips
per board (thus 1/8 processors of a full board). 
The board has basically the same function as
GRAPE-6A \citep{FMK05}, but it has an advantage that it can be
installed in 2U server unit of PC clusters due to the short height of
its heat sink. 

Figure \ref{fig:speedup} shows the speedup of the performance. 
The vertical axis denotes the clock time by single
node run (two processors) divided by the time for $N_{\rm node}$ nodes, and the horizontal axis shows the number of nodes.
The dotted line represents the case of ideal speedup, and the symbols
correspond to the runs with various number of sources ($N_{\rm source} =
2, 10, 20$).

Some of present results are slightly better than ideal case
(e.g. $N_{\rm node}=8, ~N_{\rm source} = 20$ case). 
The chief reason of this super linearity is the different convergence
speed of implicit solver for different number of nodes. 
The numerical calculations between the runs with
different number of nodes are not exactly the same ( see section \ref{RT} ).
Consequently, the number of time steps vary from runs to runs and
the convergence speed for each time step is also not identical. Thus,
such different convergence property could lead to the super linear speed
up. 
Apart from the fluctuation due to this unevenness of time steps,
parallelization efficiency is quite good as far as the number of nodes
are smaller than 16 for $N_{\rm source} \ge 10$.  

For $(N_{\rm node},N_{\rm source}) =(16, 2)$ case,
 the speedup is not as good as those for other runs. 
This run include only 2 sources, which
number is much smaller than the number of nodes. Thus the 'waiting'
time in the ray tracing routine could emerge. In fact, for larger number
 of sources, we do not observe such reduction of parallelization efficiency.
However, this result also means when we increase the number of nodes to
 $\sim 100$, similar phenomenon is expected with $N_{\rm source}\sim
 10$. 
Therefore, we have to keep in mind this property when we run the code
 with much larger
 number of nodes in the future after the completion of full FIRST cluster.

\section{Summary}
\label{summary}
We have constructed the new code for radiation hydrodynamics
designed to be applied to the feedback effects of first generation
stars/galaxies. The code can deal with multiple sources, and it can
solve the transfer of ionizing photons and photodissociating photons.

The code is already parallelized, and almost ready to be
installed in the newly developed 
full size huge PC cluster ``FIRST'' in University
of Tsukuba. We compare the numerical results with one dimensional
calculations, and find the good agreements.

\bigskip
We thank the anonymous referee for important comments.
We also thank  M. Umemura and T. Nakamoto, K. Ohsuga for careful reading of
the manuscript.  We thank N. Shibazaki for
continuous encouragement. We thank H. Maki for providing the 1D-PPM code.
We also thank T. Kitayama for providing the 1-dimensional results, and
thank all the collaborators in TSU3 project
( http://www.mpa-garching.mpg.de/tsu3 ) for fruitful discussions during
the workshop in CITA.
The analysis has been made with computational facilities 
at Center for Computational Science in University of Tsukuba and Rikkyo
University. 
This work was supported in part by Ministry of Education, Culture,
Sports, Science, and Technology (MEXT), Grants-in-Aid, Specially
Promoted Research 16002003 and Young Scientists (B) 17740110.

\appendix
\section{Volume averaged photoionization rate with helium}
\label{app}
In this section, we show the basic idea of the volume averaged
photoionization/photoheating rate for H/He gas. It
could be useful to the readers although we have not implemented yet.
First of all, we have the radiation transfer equation assuming the pure
absorption (i.e. on-the-spot approximation):
\begin{eqnarray}
\frac{dI_\nu}{dr} = - ( n_{\rm HI}\sigma_\nu^{\rm HI} + n_{\rm
 HeI}\sigma_\nu^{\rm HeI} + n_{\rm HeII}\sigma_\nu^{\rm HeII})I_\nu\label{eq:radtrall}
\end{eqnarray}
where $\sigma_\nu^{\rm X}$ and $n_{\rm X}$ denote the photoionization
cross section and number density of X species, respectively. 
The photoionization rates for HI, HeI and HeII are
\begin{eqnarray}
k_{\rm ion}^{\rm HI}&=& n_{\rm HI}\int_{\nu_L}^\infty\!\!\!\!\int
 \frac{I_\nu}{h\nu} \sigma_{\nu}^{\rm HI} d\Omega d\nu,
 \label{eq:rateionHI} \\
k_{\rm ion}^{\rm HeI}&=& n_{\rm HeI}\int_{\nu_{L{\rm HeI}}}^\infty\!\!\!\!\int
 \frac{I_\nu}{h\nu} \sigma_{\nu}^{\rm HeI} d\Omega d\nu,
 \label{eq:rateionHeI} \\
k_{\rm ion}^{\rm HeII}&=& n_{\rm HeII}\int_{\nu_{L{\rm HeII}}}^\infty\!\!\!\!\int
 \frac{I_\nu}{h\nu} \sigma_{\nu}^{\rm HeII} d\Omega d\nu,
 \label{eq:rateionHeII} 
\end{eqnarray}
Here $\nu_L$,$\nu_{L{\rm HeI}}$ and $\nu_{L{\rm HeII}}$ denote the Lyman limit frequency of HI,
HeI, HeII respectively. 

Combining equation(\ref{eq:radtrall}) with
equations(\ref{eq:rateionHI})-(\ref{eq:rateionHeII}), and perform the
integration in solid angle, we have
\begin{eqnarray}
k_{\rm ion}^{\rm HI}&=& -\frac{1}{4\pi r^2}\int_{\nu_L}^\infty \frac{n_{\rm
 HI}\sigma_\nu^{\rm HI}}{(n\sigma)_{\rm tot}}\frac{d}{dr}\frac{L_\nu^*\exp\left(-\tau_\nu\right)}{h\nu} d\nu,
\label{eq:rateion2HI}\\
k_{\rm ion}^{\rm HeI}&=& -\frac{1}{4\pi r^2}\int_{\nu_{L{\rm HeI}}}^\infty \frac{n_{\rm
 HeI}\sigma_\nu^{\rm HeI}}{(n\sigma)_{\rm tot}}\frac{d}{dr}\frac{L_\nu^*\exp\left(-\tau_\nu\right)}{h\nu} d\nu,
\label{eq:rateion2HeI}\\
k_{\rm ion}^{\rm HeII}&=& -\frac{1}{4\pi r^2}\int_{\nu_{L{\rm HeII}}}^\infty \frac{n_{\rm
 HeII}\sigma_\nu^{\rm HeII}}{(n\sigma)_{\rm tot}}\frac{d}{dr}\frac{L_\nu^*\exp\left(-\tau_\nu\right)}{h\nu} d\nu,
\label{eq:rateion2HeII}
\end{eqnarray}
where 
\begin{eqnarray}
(n\sigma)_{\rm tot}&\equiv& n_{\rm HI}\sigma_\nu^{\rm HI} + n_{\rm
 HeI}\sigma_\nu^{\rm HeI} + n_{\rm HeII}\sigma_\nu^{\rm HeII}\nonumber \\ 
\tau_\nu&\equiv& \int_0^r (n\sigma)_{\rm tot} dr \nonumber 
\end{eqnarray}

Then we assume that the ratio such as $n_{\rm X}\sigma_\nu^{\rm
X}/(n\sigma)_{\rm tot}$ in equations
(\ref{eq:rateion2HI})-(\ref{eq:rateion2HeII}) are independent of $r$ within the
single grid, which is a regular assumption for finite difference
scheme. Therefore we can regard the ratio as constants when we integrate
the equation within a single grid.
The volume averaged rates are as follows:
\begin{eqnarray}
\overline{k_{\rm ion}^{\rm HI}} &=& \frac{3}{\Delta r_i}
 \frac{\Phi_1^{\rm HI}(r_i-\Delta r_i/2)-\Phi_1^{\rm HI}(r_i + \Delta r_i/2)}{3r_i^2 + \Delta r_i^2/4},\label{eq:rateion3HI}\\
\overline{k_{\rm ion}^{\rm HeI}} &=& \frac{3}{\Delta r_i}
 \frac{\Phi_1^{\rm HeI}(r_i-\Delta r_i/2)-\Phi_1^{\rm HeI}(r_i + \Delta r_i/2)}{3r_i^2 + \Delta r_i^2/4},\label{eq:rateion3HeI}\\
\overline{k_{\rm ion}^{\rm HeII}} &=& \frac{3}{\Delta r_i}
 \frac{\Phi_1^{\rm HeII}(r_i-\Delta r_i/2)-\Phi_1^{\rm HeII}(r_i + \Delta r_i/2)}{3r_i^2 + \Delta r_i^2/4},\label{eq:rateion3HeII}
\end{eqnarray}
where
\begin{eqnarray}
\Phi_1^{\rm HI}(r)&=&\int_{\nu_L}^\infty \frac{n_{\rm
 HI}\sigma_\nu^{\rm HI}}{(n\sigma)_{\rm tot}}\frac{L_\nu^*}{4\pi}\frac{\exp\left(-\tau_\nu\right)}{h\nu} d\nu,\label{eq:numionHI}\\
\Phi_1^{\rm HeI}(r)&=&\int_{\nu_{L{\rm HeI}}}^\infty\frac{n_{\rm
 HeI}\sigma_\nu^{\rm HeI}}{(n\sigma)_{\rm tot}} \frac{L_\nu^*}{4\pi}\frac{\exp\left(-\tau_\nu\right)}{h\nu} d\nu,\label{eq:numionHeI}\\
\Phi_1^{\rm HeII}(r)&=&\int_{\nu_{L{\rm HeII}}}^\infty \frac{n_{\rm
 HeII}\sigma_\nu^{\rm HeII}}{(n\sigma)_{\rm tot}}\frac{L_\nu^*}{4\pi}\frac{\exp\left(-\tau_\nu\right)}{h\nu} d\nu.\label{eq:numionHeII}
\end{eqnarray}

The photoheating rates are evaluated by exactly the same way as
above. The volume averaged photoheating rates are
\begin{eqnarray}
\overline{\Gamma_{\rm ion}^{\rm HI}} &=& \frac{3}{\Delta r_i}
 \frac{\Phi_2^{\rm HI}(r_i-\Delta r_i/2)-\Phi_2^{\rm HI}(r_i + \Delta r_i/2)}{3r_i^2 + \Delta r_i^2/4},\label{eq:rateheat3HI}\\
\overline{\Gamma_{\rm ion}^{\rm HeI}} &=& \frac{3}{\Delta r_i}
 \frac{\Phi_2^{\rm HeI}(r_i-\Delta r_i/2)-\Phi_2^{\rm HeI}(r_i + \Delta r_i/2)}{3r_i^2 + \Delta r_i^2/4},\label{eq:rateheat3HeI}\\
\overline{\Gamma_{\rm ion}^{\rm HeII}} &=& \frac{3}{\Delta r_i}
 \frac{\Phi_2^{\rm HeII}(r_i-\Delta r_i/2)-\Phi_2^{\rm HeII}(r_i + \Delta r_i/2)}{3r_i^2 + \Delta r_i^2/4},\label{eq:rateheat3HeII}
\end{eqnarray}
where
\begin{eqnarray}
\Phi_2^{\rm HI}(r)&=&\int_{\nu_L}^\infty\frac{n_{\rm
 HI}\sigma_\nu^{\rm HI}}{(n\sigma)_{\rm tot}}
 \frac{L_\nu^*}{4\pi}\frac{\exp\left(-\tau_\nu\right)}
{h\nu}\left(h\nu - h\nu_L\right) d\nu,\label{eq:numheatHI}\\
\Phi_2^{\rm HeI}(r)&=&\int_{\nu_{L{\rm HeI}}}^\infty\frac{n_{\rm
 HeI}\sigma_\nu^{\rm HeI}}{(n\sigma)_{\rm tot}}
 \frac{L_\nu^*}{4\pi}\frac{\exp\left(-\tau_\nu\right)}{h\nu} 
\left(h\nu - h\nu_{L{\rm HeI}}\right) d\nu,\label{eq:numheatHeI}\\
\Phi_2^{\rm HeII}(r)&=&\int_{\nu_{L{\rm HeII}}}^\infty\frac{n_{\rm
 HeII}\sigma_\nu^{\rm HeII}}{(n\sigma)_{\rm tot}}
 \frac{L_\nu^*}{4\pi}\frac{\exp\left(-\tau_\nu\right)}{h\nu}
\left(h\nu - h\nu_{L{\rm HeII}}\right) d\nu.\label{eq:numheatHeII}
\end{eqnarray}

It is also worth to point out that the functions $\Phi_1^{\rm X}$ and $\Phi_2^{\rm X}$ can
be assessed only by the optical depth at three frequencies
($\nu_L$, $\nu_{L{\rm HeI}}$ and $\nu_{L{\rm HeII}}$), although the
expressions (\ref{eq:numionHI}) - (\ref{eq:numionHeII}) and
(\ref{eq:numheatHI}) - (\ref{eq:numheatHeII}) include frequency
dependent optical depth
$\tau_\nu$ \citep{NUS01}. Consequently, we  can create the tables of
ionization rates and heating rates as functions of optical depth and the
ratio $n_{\rm X}\sigma_{\nu}^{\rm X}/(n\sigma)_{\rm tot}$ at
Lyman limits ($\nu_L$, $\nu_{L{\rm HeI}}$ and $\nu_{L{\rm HeII}}$).




\setcounter{figure}{0}

\clearpage
\begin{figure}
\begin{center}
\FigureFile(12cm,7cm){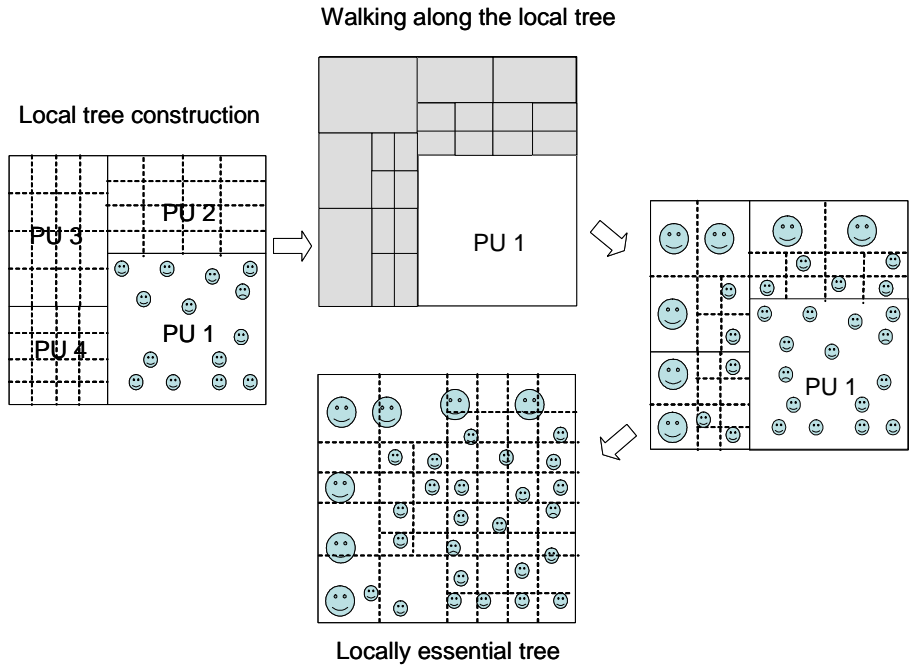} 
\caption{Parallel tree scheme to assess the force of gravity on the
 particles included in a processor(PU1) is schematically shown.
Left panel shows the initial phase of the
 algorithm, in which local tree is made on each processor. 
In the next
 step, the tree nodes which can interact the particles in PU1 are found
 by walking along the local tree. The data of the nodes are sent to PU1
 as ``particles'' that do not have inner structure (upper middle panel).
 Finally, we have the complete list of particles which can interact with the
 particles in PU1(right panel). Utilizing this particle list, we can
 assess the force of gravity on the  particles in PU1
 by building up and walking down the locally essential
 tree (lower middle).
}\label{fig:Tree}
\end{center}
\end{figure}

\begin{figure}
\begin{center}
\FigureFile(12cm,7cm){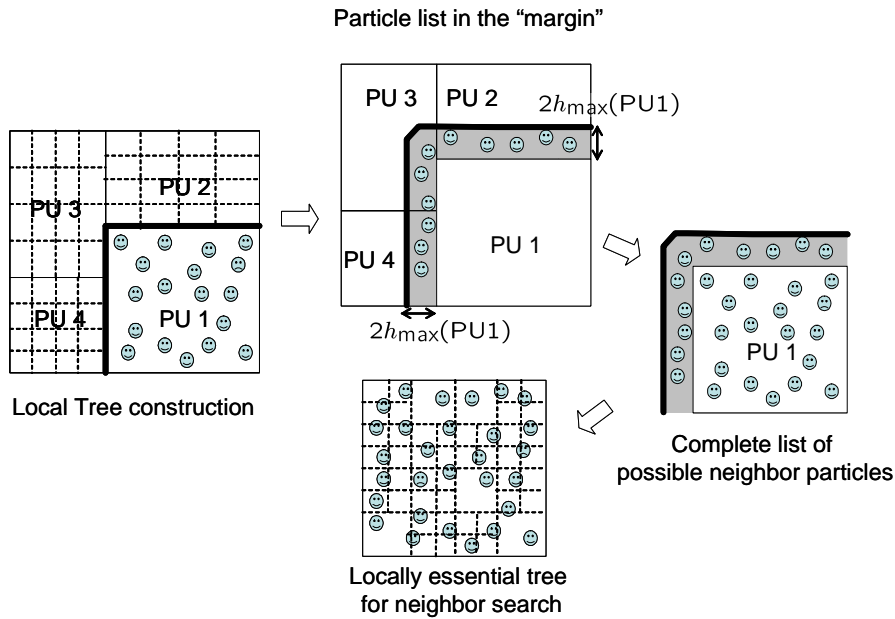} 
\caption{Employed neighbor search algorithm on parallel computer system
 to obtain the neighbour list of a certain processor (PU1) is shown.
Left panel shows the initial phase of the
 algorithm, in which local tree is made on each processor. In the next
 step, the data of particles in the ``margin'' of PU1 are sent to
 PU1 (upper middle). 
Finally, the list of particles which can be in the neighbor of
 particles in PU1 is completed (right panel). Using this list, locally
 essential tree is re-constructed, which is used to find neighbor
 particles in PU1 ( lower middle ).
}\label{fig:neighbor}
\end{center}
\end{figure}

\begin{figure}
\begin{center}
\FigureFile(12cm,8cm){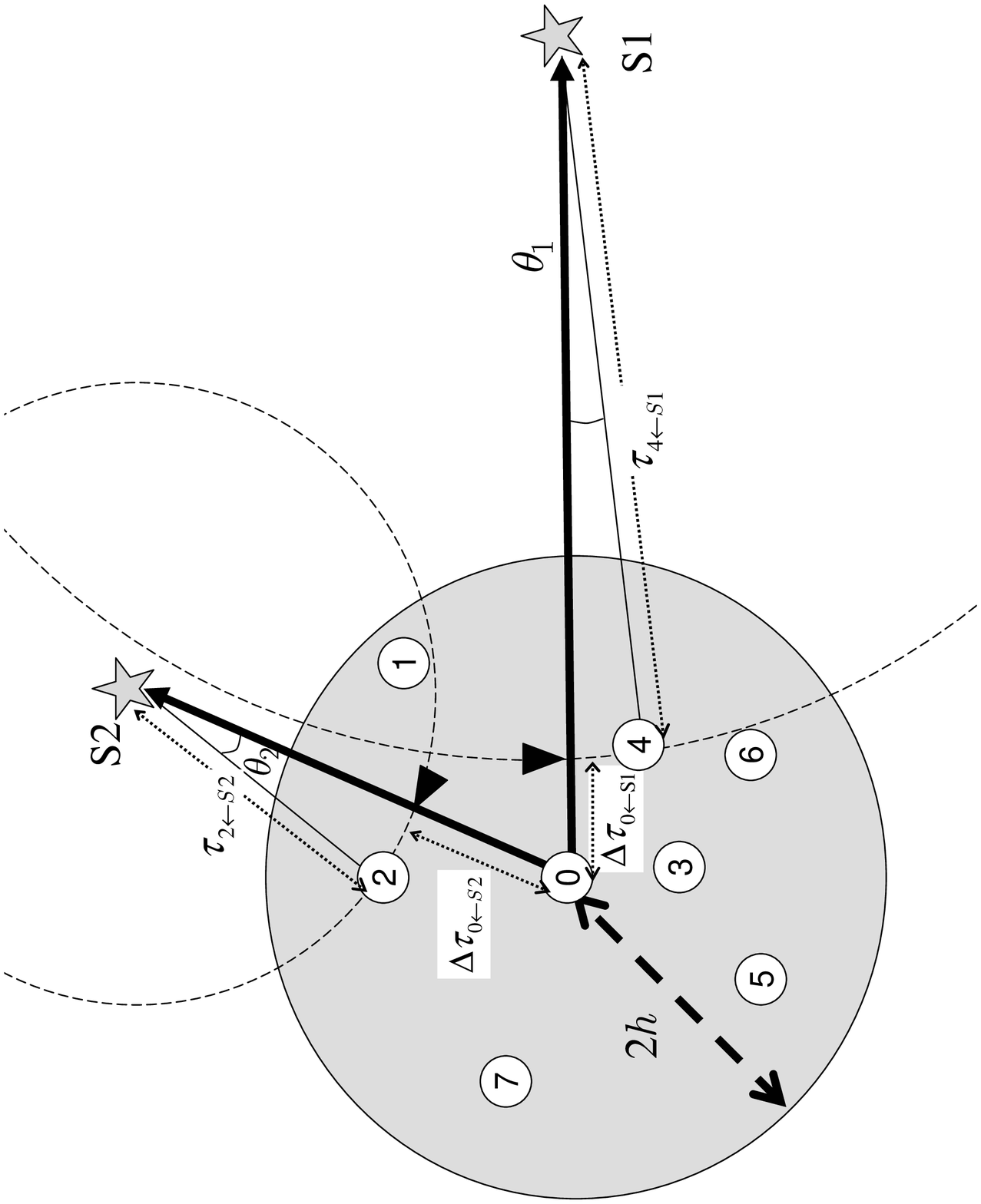}
\caption{Ray tracing method for multiple sources are
 schematically shown. Numbers in circles represent the ID of SPH
 particles in the neighbor of SPH particle ``0''. S1 and S2 represent
 the position of two sources. Two wedge-shapes show the position on
 which the optical depth from the sources are projected from
 corresponding particles (in this case 2 and 4).
}\label{fig1}
\end{center}
\end{figure}
\clearpage

\begin{figure}
\begin{center}
\FigureFile(12cm,7cm){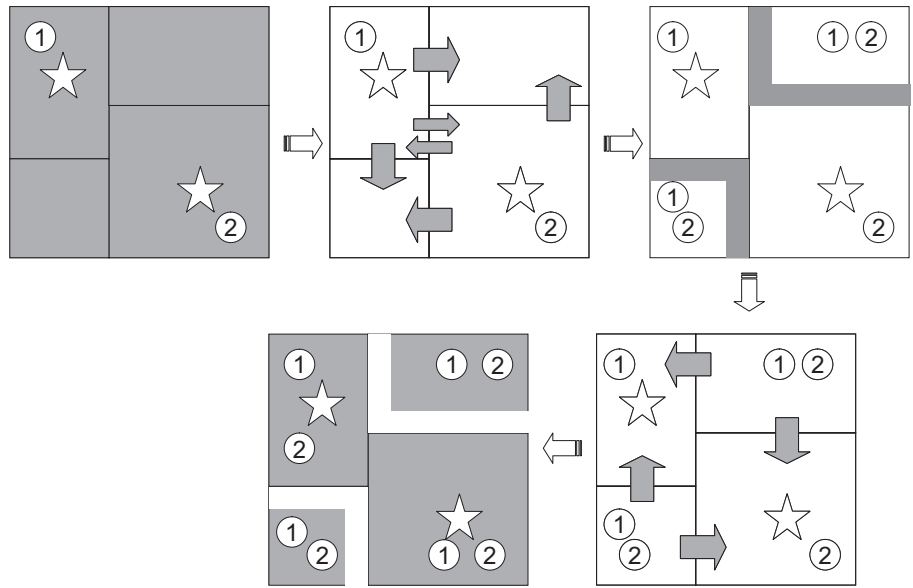}
\caption{Order of computations and communications for ray tracing
 on parallel machine is shown schematically for an example. In this
 example, the computational domain is decomposed into 2D, and consists of 4
 nodes. The calculations and communications proceed along the direction
 of small open arrows. The position of two sources are shown by star-shaped
 symbols. The shaded region on the nodes are
 occupied by computation, and the shaded arrows represents the
 direction of the present communications.
}\label{fig2}
\end{center}
\end{figure}

\begin{figure}
\begin{center}
\FigureFile(15cm,15cm){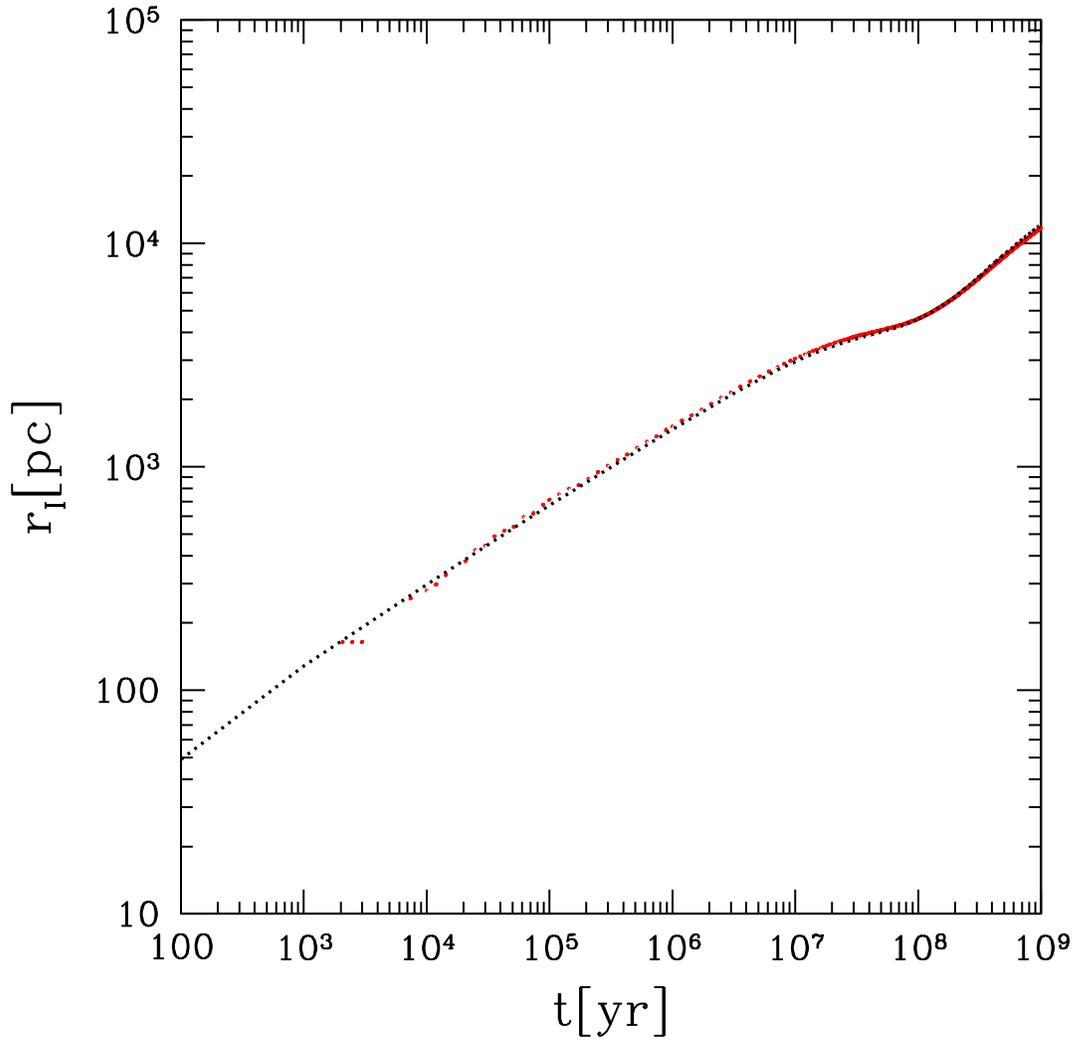}
\caption{Propagation of ionization front in uniform media is
 shown. Initial density and temperature 
is $n_{\rm H} = 0.01 {\rm cm^{-3}}, T=3\times 10^2{\rm K}$.
The ionizing photon luminosity of the source is $S=1.33\times 10^{50} {\rm
s^{-1}}$ and the spectrum is black
body with $T_*= 9.92\times 10^4$K.
 Horizontal axis shows the time after the ignition of central
 star, and the vertical axis shows the position of ionization front, defined as the
 radius at which the fraction of HI is 0.5. Red dots denote the results
 from the SPH code, whereas the dotted line denotes the ionization front
 position from the 1D simulation.
}\label{fig:w0if}
\end{center}
\end{figure}

\begin{figure}
\begin{center}
\FigureFile(15cm,15cm){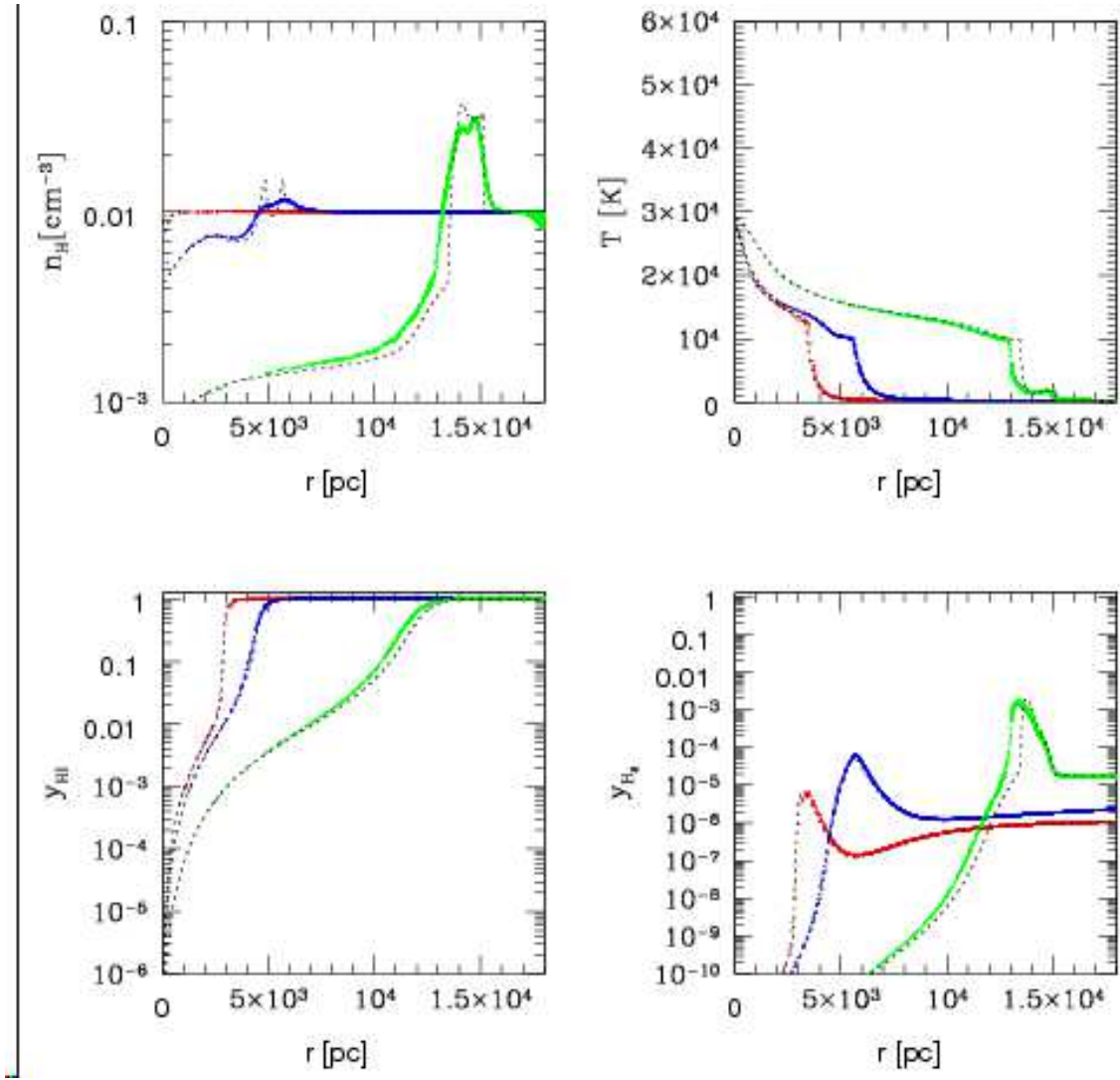}
\caption{Spatial distribution of physical quantities at three
 time snapshots ($t=10^7,10^8,10^9$yr) are
 shown for the run shown in Fig.\ref{fig:w0if}.
Upper panels show the density and temperature, and the lower
 panels show the HI fraction and H$_2$ fraction. Horizontal axes denote
 the distance from the source. The dotted lines denote the results from
 the 1D simulation.
}\label{fig:w0}
\end{center}
\end{figure}

\begin{figure}
\begin{center}
\FigureFile(15cm,15cm){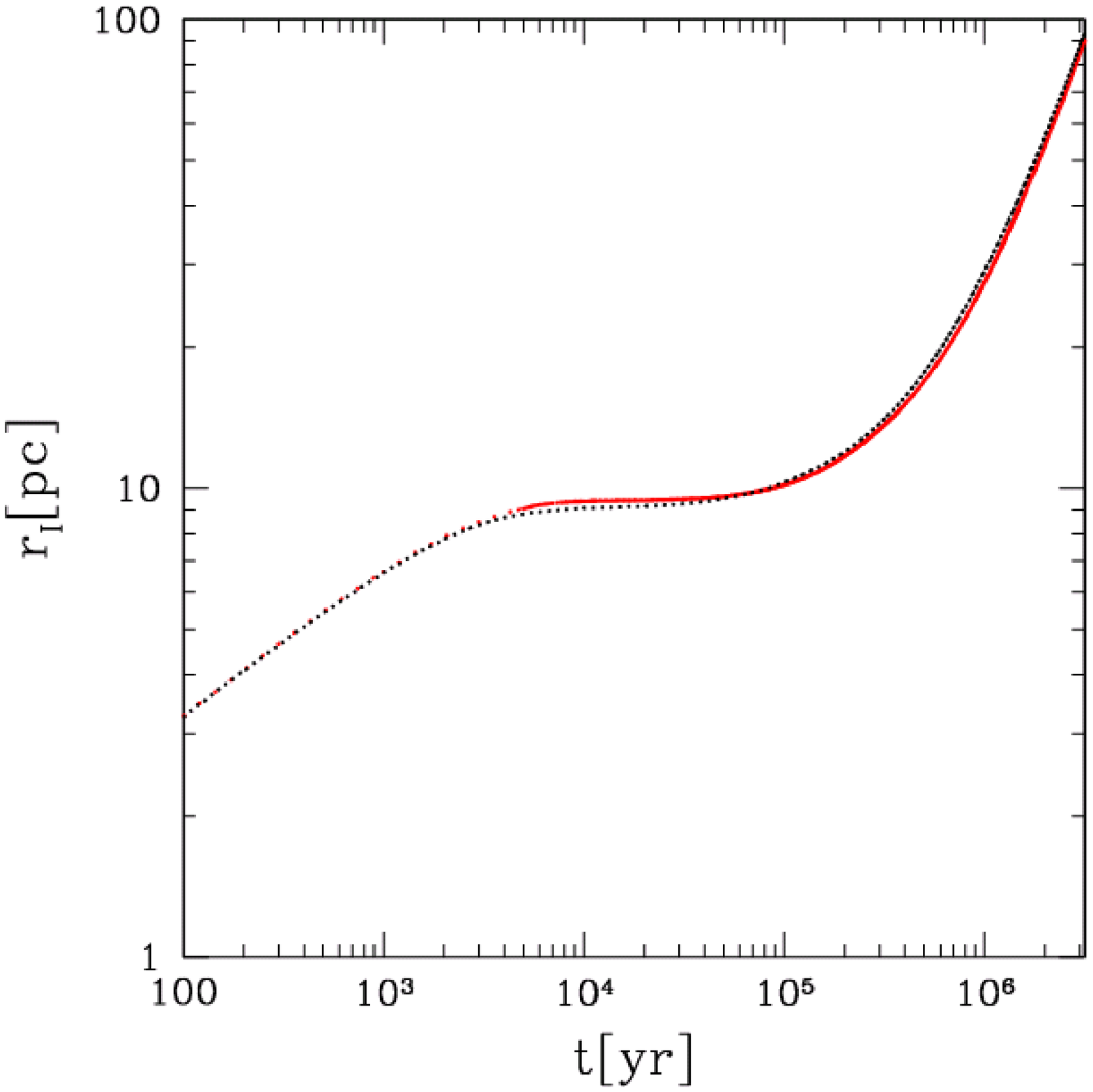}
\caption{Same as Fig.\ref{fig:w0if}, except that the initial
 density distribution is core-envelope structure, discussed in section\ref{coreenv}.
}\label{fig:w2if}
\end{center}
\end{figure}

\begin{figure}
\begin{center}
\FigureFile(15cm,15cm){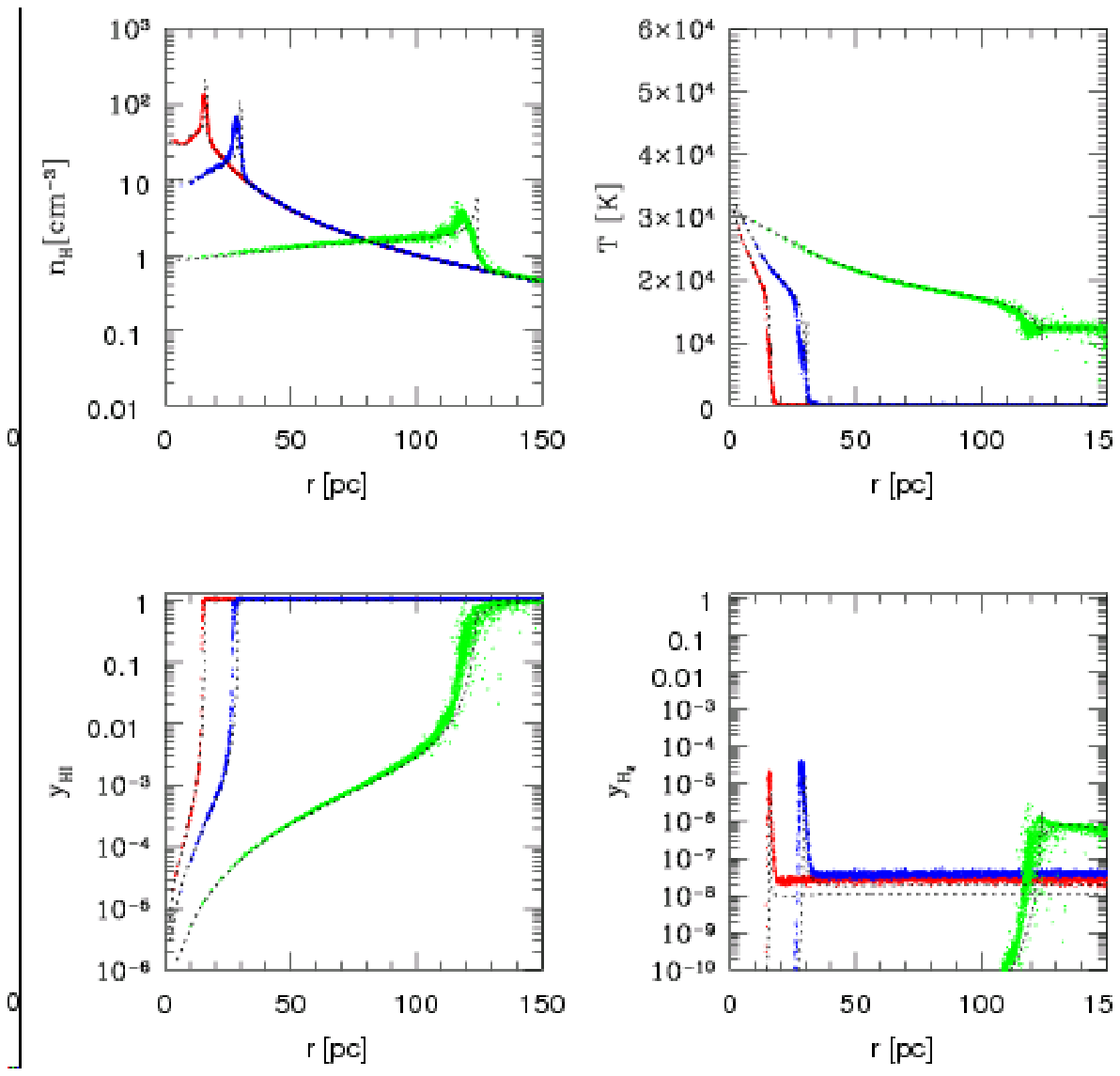}
\caption{Same as Fig.\ref{fig:w0}, except that the initial
 density distribution is core-envelope structure, discussed in section\ref{coreenv}
}\label{fig:w2}
\end{center}
\end{figure}

\begin{figure}
\begin{center}
\FigureFile(12cm,8cm){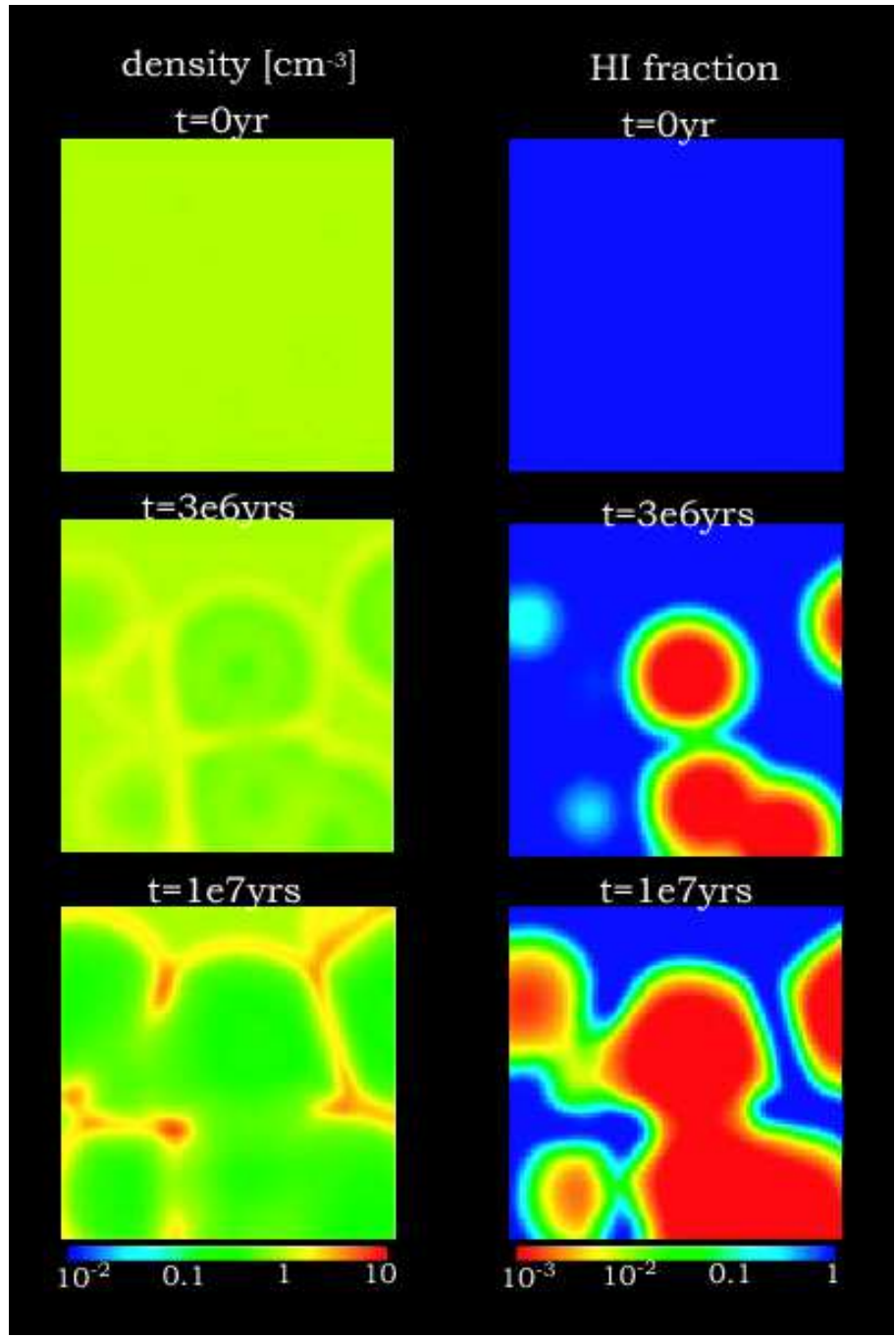}
\caption{Spatial slice of the multi-source run in section
 \ref{multi_source} is shown.  Left panels show the evolution of
 density, and the ionization structures are shown in the right
 panels. The color legends are shown at the bottom.
}\label{fig:multi_test}
\end{center}
\end{figure}

\begin{figure}
\begin{center}
\FigureFile(15cm,15cm){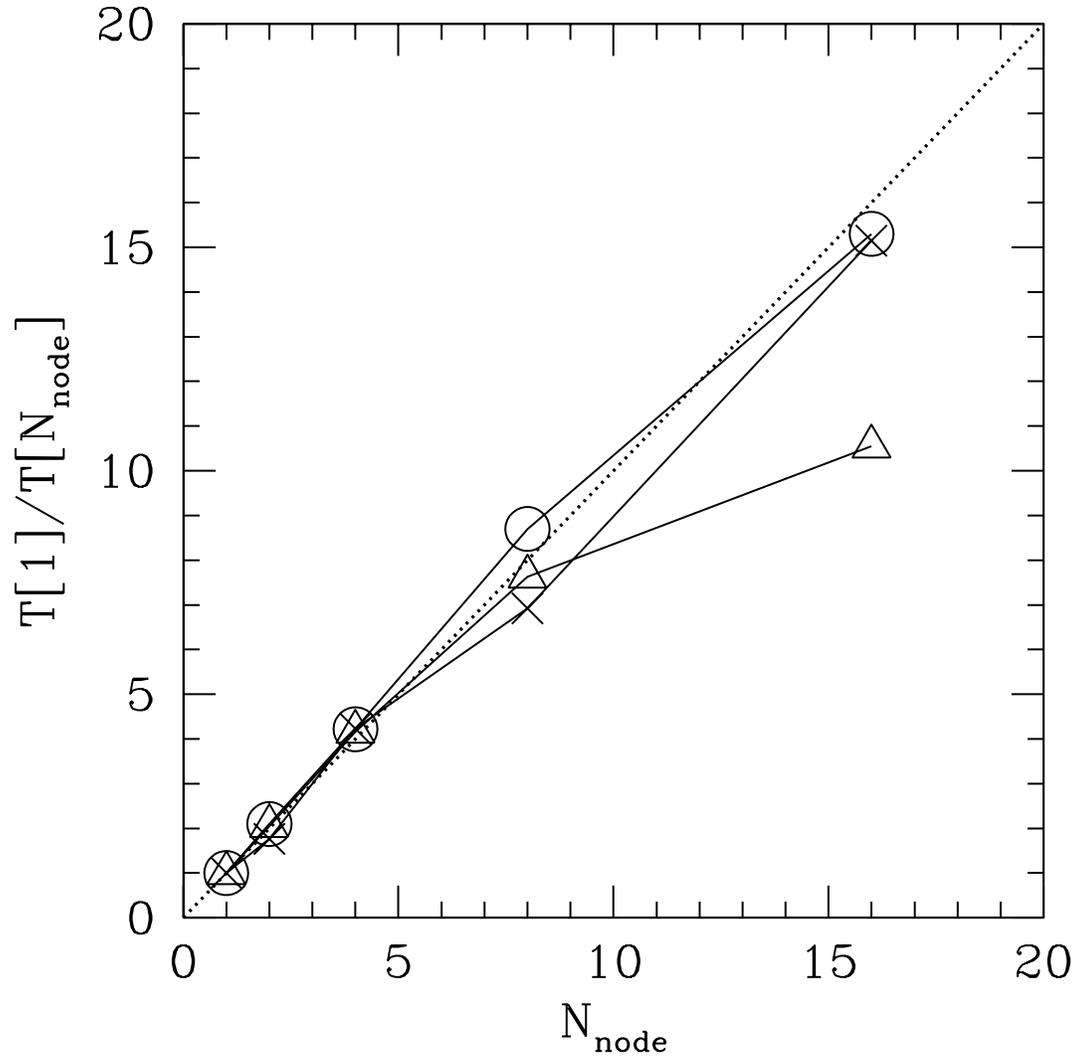}
\caption{Speedup of calculations by parallelization is shown.
Vertical axis denotes the clock time for one node (two processors)
 divided by the clock time with $N_{\rm node}$ nodes. Horizontal axis shows
 the number of nodes. The dotted line represents the case of ideal speedup.
 Symbols connected by the solid lines are the results with different number of sources ($N_{\rm
 source}$): open circles show the results for $N_{\rm source} = 20$,
 vertices for $N_{\rm source} = 10$, and triangles for $N_{\rm source} = 2$.    
}\label{fig:speedup}
\end{center}
\end{figure}

\end{document}